\documentclass[aps,prl,showpacs,twocolumn]{revtex4}
\usepackage{amsmath, amssymb}

\newcommand{\sign}{\mbox{sign }}
\newcommand{\even}{\mbox{ even}}
\newcommand{\odd}{\mbox{ odd}}
\newcommand{\Sd}{\mbox{$S_d$}}
\newcommand{\ket}[1]{\mbox{$|#1\rangle$}}
\newcommand{\bra}[1]{\mbox{$\langle#1|$}}
\newcommand{\otimesd}{\mbox{${}^{\otimes d}$}}
\newcommand{\PT}{\mbox{${}^{\rm T_2}$}}

\begin{document}
\title{Non-existence of bipartite bound entanglement with negative partial transposition}

\author{J. Sperling} \email{jan.sperling2@uni-rostock.de}
\author{W. Vogel} \email{werner.vogel@uni-rostock.de}
\affiliation{Arbeitsgruppe Quantenoptik, Institut f\"ur Physik, Universit\"at Rostock, D-18051 Rostock, Germany}


\begin{abstract}
Bound entanglement with a nonpositive partial transposition (NPT) does not exist.
For any NPT entangled state a distillation procedure can be based on a certain number of copies.
This number is the minimal Schmidt rank of a pure state needed to witness the NPT entanglement under study.
\end{abstract}

\date{\today}

\maketitle


Entanglement plays a major role for the vast fields of Quantum Information and Quantum Technology, for an introduction see~\cite{book-chuang,book-horodecki,book-guehne}.
First, the phenomenon entanglement was studied in~\cite{epr,schroedinger}.
A precise definition of the notion of entanglement was given in~\cite{werner}. An important method for identifying entanglement was given in term of the partial transposition (PT)~\cite{peres}.
A quantum state is entangled, if it exhibits negativities under PT.

In general, the PT condition for entanglement is only a necessary one.
It is also sufficient for the case of a $2\otimes 2$ or $2\otimes 3$ Hilbert space~\cite{horodecki3}, and for the case of Gaussian states~\cite{gaussian1,gaussian2}.
The general identification of continuous variable quantum states with negativities under partial transposition (NPT) has been given in~\cite{eigen-shchukin}. 
There exist, however, entangled quantum states with a positive semi-definite partial transposition (PPT)~\cite{horodecki2,cvbe}.
The general problem of identifying entangled states has been studied in~\cite{doherty,eigen-sperling1}, which eventually allows to identify PPT states.
It is well known that PPT states cannot be used for the purification or distillation of a Bell-type state~\cite{horodecki4}.

In general, quantum states which are not distillable are called bound entangled (BE) states, for an introduction see~\cite{book-guehne,book-horodecki,phd-clarisse}.
It was shown that in the case of a $2\otimes d$ bipartite quantum system all NPT states are distillable~\cite{duer2}.
For the general case of a bipartite quantum system, the validity of the conjecture that {\em bound entanglement is equivalent to entanglement and positivity of the partial transposition} -- or alternatively: all NPT states are distillable -- remained an open problem.
For some considerations of this problem we refer to~\cite{horodecki1,duer1,divincenzo}.

We would like to emphasize that the solution of this problem is not of academic interest only~\cite{book-horodecki}, there is
''{\em the long-standing and still open question of the existence of bound entangled states violating the positive partial transpose criterion, which would have severe consequences for communication theory.}''
The answer to this question has fundamental implications for the properties of distillable entangled states:
''{\em The problem of existence of NPT BE states has important consequences.
If they indeed exist, then distillable entanglement is nonadditive and nonconvex. \dots
The set of bipartite BE states will not be closed under tensor product, and under mixing.}''

In the present contribution we consider the distillation of an arbitrary bipartite NPT quantum state. We show that any state of this type can be distilled.
This proves the conjecture that all bipartite BE states are PPT and entangled.

Without loss of generality we consider a finite Hilbert space, $\mathcal H=\mathcal H_1\otimes\mathcal H_2$.
The generalization to continuous-variable systems follows immediately~\cite{eigen-sperling2}.
Further on we may assume in the following that the PT map acts on the second system, $\rm T_2$.
The NPT condition for entanglement reads as~\cite{peres}
\begin{align}
	\exists|\phi\rangle\in\mathcal H:\,\langle\phi|\rho^{\rm T_2}|\phi\rangle<0.
\end{align}
Negativities can also be found in terms of principal minors~\cite{eigen-shchukin}.
A unification of both ideas, together with Ref.~\cite{horodecki4}, will deliver the possibility to distill all NPT quantum states.

First of all let us define some general notions.
The permutation group is defined as
\begin{align}
	\Sd=\{f:\{1,\dots,d\}\to\{1,\dots,d\}|f\mbox{ bijective}\}.
\end{align}
The identity is denoted by $e\in\Sd$, and $\tau\in\Sd$ be an arbitrary transposition (permutation of two elements).
On this group the sign-function is given by
\begin{align}
	\sign f=\left\lbrace\begin{array}{ccc}
		+1 & \mbox{ for } & f\even\\
		-1 & \mbox{ for } & f\odd
	\end{array}\right..
\end{align}
The value $\sign f=+1$ denotes a decomposition of $f$ into an even number of transpositions, $f=\tau_1\circ\dots\circ\tau_{2n}$, and $\sign f=-1$ for an odd number of transpositions, $f=\tau_1\circ\dots\circ\tau_{2n+1}$.
Note that $\sign f=\sign f^{-1}$ and $\sign f\circ g=\sign f\cdot\sign g$.
The alternating group $A_d=\{h\in\Sd:h\even\}$ is a normal subgroup of $\Sd$, $A_d\triangleleft\Sd$ $\Leftrightarrow$ $\forall f\in\Sd:A_d=f^{-1}\circ A_d\circ f$.

For an efficient treatment of minors it is advantageous to consider the vector of a permutation $f$,
\begin{align}
	\ket{f}=&\ket{f(1)}\otimes\dots\otimes\ket{f(d)}=\ket{f(1),\dots,f(d)},
	\intertext{the identity vector}
	\ket{e}=&\ket{1}\otimes\dots\otimes\ket{d},
	\intertext{and the composition of two subsequent permutations}
	f\ket{g}=&\ket{f\circ g}=\ket{(f\circ g)(1),\dots,(f\circ g)(d)}.
\end{align}
The operation $f$ can be interpreted as the permutation matrix $P_f\otimes\dots\otimes P_f$, with $P_f\ket k=\ket{f(k)}$.

Now let us consider some properties of principal minors.
Therefore the linear map $C$ ($C:\mathbb C^d\to\mathbb C^d$), the number of copies $C\otimesd=\bigotimes_{t=1}^d C$, and the matrix elements $C_{t,s}$ are given.
The Leibniz formula reads as
\begin{align}
	\det C=&\sum_{f\in\Sd}\sign f\prod_{t=1}^d C_{t,f(t)}.
	\intertext{Using the above representation of the permutation vector we may write}
	\det C=&\bra e C\otimesd\sum_{f\in\Sd}\sign f \ket f.\label{Eq:Leibnitz}
\end{align}
It is useful to rewrite the determinant as a bilinear expression.
This ansatz will finally deliver a connection between quantum expectation values and principal minors,
\begin{align}
	\nonumber \left\langle C\otimesd\right\rangle=&\left(\sum_{g\in\Sd}\sign g \bra g\right) C\otimesd\left(\sum_{f\in\Sd}\sign f \ket f\right)\\
	=&\sum_{g,f\in\Sd}\sign(f\circ g^{-1})\underbrace{\prod_{t=1}^dC_{g(t),f(t)}}_{=\prod_{t=1}^dC_{t,(f\circ g^{-1})(t)}},\label{Eq:Cal1}
	\intertext{by substituting $h=f\circ g^{-1}$ we obtain from Eq.~(\ref{Eq:Leibnitz})}
	\left\langle C\otimesd\right\rangle=&\sum_{g\in\Sd}\sum_{h\in\Sd}\sign h \bra e C\otimesd \ket h=|\Sd|\cdot\det C.\label{Eq:Cal2}
\end{align}
Further on we can consider the following superpositions:
\begin{align}
	\ket \psi=&\sum_{f\in\Sd}\sign f\ket f=\ket + -\ket -.
\end{align}
Together with
\begin{align}
	\ket +=\sum_{f\even} \ket f, \quad&\quad \ket -=\sum_{f\odd} \ket f,
\end{align}
we obtain
\begin{align}
	|\Sd|\cdot\det C=\left\langle C\otimesd\right\rangle=\bra \psi C\otimesd \ket \psi.\label{Eq:BiDet}
\end{align}
Note that $\det AB=\det A\cdot \det B$.

In the following let us consider a state $\ket \phi$ which delivers $\bra \phi \rho\PT\ket \phi<0$.
The Schmidt decomposition~\cite{book-chuang} reads as $\ket \phi=\sum_{k=1}^d \lambda_k \ket{k,k}$.
Let us assume that the Schmidt rank $d$ of $\ket \phi$ is the smallest integer, which delivers negativities.
Obviously the minor with $(k,k,l,l)$-elements is negative and it has the smallest number of elements.
The matrix $C$ is given by $C_{i,j,k,l}=\bra{i,j}\rho\PT\ket{k,l}$, with $i.j,k,l\in\{1,\dots,d\}$.
The minor, resulting from the state $|\phi\rangle$, with the $(k,k,l,l)$-elements is denoted as $\det_\phi$.
Now we obtain, in analogy to Eqs.~(\ref{Eq:Leibnitz})~and~(\ref{Eq:BiDet}),
\begin{align}
	\nonumber \det{}_\phi C&=\sum_{f\in\Sd}\sign f\bra{e,e}C\otimesd\ket{f,f}\\
	&=\frac{1}{|\Sd|}\bra{\tilde \psi}C\otimesd\ket{\tilde \psi}<0,
\end{align}
with
\begin{align}
	\nonumber \ket{\tilde \psi}&=\sum_{f\in\Sd} \sign f\ket{f,f}\\
	&=\sum_{f\even}\ket{f,f}-\sum_{f\odd}\ket{f,f}.
\end{align}
Our desired two qubit state is
\begin{align}
	\ket \psi&=\ket{+,+}-\ket{-,-}\\
	&=\sum_{f,g\even}\ket{f,g}-\sum_{f',g'\odd}\ket{f',g'}.
	\intertext{The substitutions $g'=g\circ\tau$, $f'=f\circ\tau$ deliver}
	\ket \psi&=\left(\sum_{f,g\even} f\otimes g\right)(\ket{e,e}-\ket{\tau,\tau}).
\end{align}
Together with the substitution $h\circ f=g$ ($h\even$), the prefactor can be written as
\begin{align}
	\sum_{f,g\even} f\otimes g&= \sum_{h,f\even} (e\circ f)\otimes (h\circ f)\\
	\nonumber &=\left(e\otimes\sum_{h \even} h\right)\circ \left(\sum_{f\even} f\otimes f\right).
\end{align}
Thus we obtain,
\begin{align}
	\ket \psi=\left(e\otimes\sum_{h \even} h\right)\ket{\tilde\psi}.
\end{align}
The state $\ket \psi$ is a local transformation of $\ket{\tilde\psi}$.

Now let us calculate, similar as in Eqs.~(\ref{Eq:Cal1})~and~(\ref{Eq:Cal2}), our desired expectation value:
\begin{align}
	\label{Eq:ExpVal}\langle\psi|C\otimesd|\psi\rangle=&|\Sd|\cdot|A_d|\\
	\nonumber&\times\sum_{h\even}\sum_{f\in\Sd}\sign f\bra{e,e}C\otimesd\ket{f,h\circ f}.
\end{align}
Let us consider one term of Eq.~(\ref{Eq:ExpVal}),
\begin{align}
	\nonumber &\sum_{f\in\Sd}\sign f\bra{e,e}C\otimesd\ket{f,h\circ f}\\
	\nonumber =&\sum_{f\in\Sd}\sign f\bra{e,e}(C[\mathbb I\otimes P_h])\otimesd\ket{f,f}\\
	=&\det{}_{\phi} (C[\mathbb I\otimes P_h])=\det{}_\phi{C}\cdot\lambda,
\end{align}
with $\lambda=\det{}_{\phi}[\mathbb I\otimes P_h]$.
Explicitly, $\lambda$ is given by
\begin{align}
	\nonumber \lambda=&\sum_{f\in\Sd}\sign f \bra{e,e}(\mathbb I\otimes P_h)\otimesd\ket{f,f}\\
	=&\bra e P_h\otimesd \ket e=\langle e|h\rangle=\delta_{e,h}.\label{Eq:Term}
\end{align}

Thus we obtain from Eqs.~(\ref{Eq:ExpVal})~and~(\ref{Eq:Term}),
\begin{align}
	\bra{\psi}C\otimesd\ket{\psi}=|A_d|\cdot \bra{\tilde\psi}C\otimesd\ket{\tilde\psi}<0.
\end{align}
Therefore it exists a Schmidt rank two state $\ket{\psi}$, with $\bra{\psi}(\rho\PT)\otimesd\ket{\psi}<0$.
Hence $\rho$ is destillable, according to the results in Ref.~\cite{horodecki4}.
The distillation needs $d$ copies, since $d$ is the smallest Schmidt rank of a state exhibiting negativities.


In conclusion we have shown, that all NPT states are distillable.
Together with the well known fact that entangled PPT states are bound entangled, the conjectures {\em bound entangled states are PPT states}, 
or equivalently, {\em NPT states are always distillable}, haven been proven to be correct.
The number of copies needed for the destillation is given by the minimal Schmidt rank among all the pure states witnessing the NPT entanglement.

\section*{Acknowledgment}
We greatfully acknowledge valuable comments by
F.~Brandao,
J.~Calsamiglia,
L.~Chen,
M.~Christandl,
R.~Duan,
J.~Eisert
A.~Harrow,
M.~Huber,
P.~Hyllus,
N.~Johnston,
T.~Kiesel,
O.~Oreshkov,
A.~Osterloh,
M.~Piani,
V.~Vedral,
J.~di~Vinciente,
K.~Vollbrecht,
A.~Winter,
and
M.~Wolf.

This work was supported by the Deutsche Forschungsgemeinschaft through SFB 652.

\end{document}